\documentclass[12pt]{elsarticle}                                          
\usepackage{axodraw4j}
\usepackage{hyperref}
\usepackage{graphicx}                                                         
\usepackage{a41}                                                                
\usepackage{xcolor}                     
\usepackage{slashed}
\usepackage[rflt]{floatflt}
\usepackage{float}
\usepackage{hyperref}
\hypersetup{colorlinks=true, 
linkcolor=blue, filecolor=blue, urlcolor=mygreen}
\usepackage{breakurl} 
\usepackage{times}     %
\usepackage{array}
\usepackage[english]{babel}
\usepackage[latin1]{inputenc}
\usepackage[T1]{fontenc}
\usepackage{ae}
\usepackage{url}
\usepackage{amsmath, amsthm, amssymb}
\usepackage{slashed}

\usepackage{rotating}
\usepackage{graphicx}
\usepackage{comment}
\newcounter{mmacnt}
\def\restartmma{\setcounter{mmacnt}{0}}
\restartmma \catcode`|=\active
\def|#1|{\mathrm{#1}}
\catcode`|=12
\newenvironment{mma}{
\par\smallskip
\catcode`|=\active
\parskip=0pt\parindent=0pt 
\small
\def\In##1\\{%
\def\linebreak{\hfill\break\null\qquad}%
\refstepcounter{mmacnt}
\hangindent=2.5em\hangafter=0
\leavevmode
\llap{\tiny\sffamily In[\arabic{mmacnt}]:=\kern.5em}%
\mathversion{bold}\footnotesize$
\displaystyle##1$\normalsize
\mathversion{normal}\par
 }%
\def\Print##1\\{%
\def\linebreak{\hfill\break}%
\hangindent=2.5em\hangafter=0
\leavevmode ##1\par}%
\def\Out##1\\{%
\def\linebreak{$\hfill\break\null\hfill$}%
\kern\abovedisplayskip\par
\hangindent=2.5em\hangafter=0
\leavevmode
\llap{\tiny\sffamily Out[\arabic{mmacnt}]=\kern.5em}
\footnotesize$\displaystyle##1$
\normalsize\hfill\null\par
\kern\belowdisplayskip
}%
\def\Warning##1##2\\{%
\def\linebreak{\hfill\break}%
\hangindent=2.5em\hangafter=0
\leavevmode
{\scriptsize##1 : ##2}\par}%
}{%
\par\smallskip
}

\usepackage{color}
\newenvironment{fshaded}{%
\MakeFramed {\FrameRestore}
}%
{\endMakeFramed}

\newcommand{\n}{\nonumber}
\makeatletter
\def\ps@pprintTitle{%
\let\@oddhead\@empty
\let\@evenhead\@empty
\def\@oddfoot{\reset@font\hfil\thepage\hfil}
\let\@evenfoot\@oddfoot
}
\makeatother
\begin{document}                        
\begin{frontmatter}
\title{\huge 
\textbf{One-loop form factors for $H\rightarrow  
\gamma^*\gamma^*$ \\ in $R_{\xi}$ gauge}}

\author{Khiem Hong Phan}
\ead{phanhongkhiem@duytan.edu.vn}
\address{\it Institute of Fundamental and Applied Sciences, 
Duy Tan University, Ho Chi Minh City $700000$, Vietnam\\ 
Faculty of Natural Sciences, Duy Tan University, 
Da Nang City $550000$, Vietnam}
\author{Dzung Tri Tran}
\address{\it 
University of Science Ho Chi Minh City, $227$ 
Nguyen Van Cu, District $5$, HCM City, Vietnam \\
Vietnam National University Ho Chi Minh City, Linh Trung Ward, 
Thu Duc District, HCM City, Vietnam}
\pagestyle{myheadings}
\markright{}
\begin{abstract} 
In this paper, we present general one-loop form factors 
for $H\rightarrow \gamma^* \gamma^*$ in $R_{\xi}$ gauge, 
considering all cases of two on-shell, one on-shell and 
two off-shell for final photons. The calculations are 
performed in standard model and in arbitrary beyond the 
standard models which charged scalar particles may be 
exchanged in one-loop diagrams. Analytic results for 
the form factors are shown in general forms which are 
expressed in terms of the Passarino-Veltman functions. 
We also confirm the results in previous computations 
which are available for the case of two on-shell photons. 
The $\xi$-independent of the result is also discussed. 
We find that numerical results are good stability with 
varying $\xi=0,1$ and $\xi\rightarrow \infty$. 
\end{abstract}
\end{frontmatter}
\section{Introduction}
One of the main targets at future colliders such as high luminosity 
the Large Hadron Collider (HL-LHC)~\cite{ATLAS:2013hta,CMS:2013xfa}
and future lepton colliders~\cite{Baer:2013cma} is to 
measure the properties of the Standard Model Higgs boson ($H$) 
precisely. All the Higgs decay modes, Higgs boson productions 
and the couplings of Higgs to fermions, gauge bosons are measured 
precisely. From these activities, one may explore the nature 
of the Higgs sector as well as search for new physics.

Among Higgs decay modes,the decay of Higgs boson into two photons is 
the most important for several following reasons. First, this 
arises at first at one-loop diagrams. Therefore, it is 
sensitive with new physics which the charged scalar particles 
may exchange in the loop diagrams. As a result, the calculations 
for one-loop and higher-loop contributions to the 
decay amplitudes of $H\rightarrow \gamma\gamma$ play a key role in 
controlling the standard model background, constraining 
new physics parameters.  Secondly,
one-loop form factors for $H\rightarrow\gamma\gamma^*, \gamma^*\gamma^*$
($\gamma^*$ presents for a virtual photon) are useful for studying 
Higgs productions and its properties at $\gamma\gamma, e\gamma$ colliders
\cite{Muhlleitner:2005ne,Watanabe:2013ria,Watanabe:2014xaa, 
Melles:1999xd,Niezurawski:2002jx,Godbole:2002qu,Rizzo:2000kq}. 
Last but not least, the decay processes 
$H\rightarrow\gamma^* \gamma\rightarrow f\bar{f}\gamma,
\gamma^*\gamma^* \rightarrow 4$ fermions provide a 
crucial tool for controlling background for 
$ H\rightarrow f\bar{f}\gamma, H\rightarrow 4$ fermions 
at future colliders.

Many calculations for one-loop contributions to
$H\rightarrow\gamma \gamma$ within standard model (SM) and its extensions 
have been presented in~\cite{Resnick:1973vg, Shifman:1979eb,Gastmans:2011ks,Gastmans:2011wh,Wu:2016nqf,Shifman:2011ri,Huang:2011yf,Marciano:2011gm, Jegerlehner:2011jm, Shao:2011wx,Donati:2013iya,Christova:2014mea,Kile:2016ipo,Li:2017hnv,Melnikov:2016nvo}, 
also in the references therein. More recently, the authors of Ref.~\cite{Wu:2017rxt} 
argue that one-loop W boson contributions to
$H\rightarrow\gamma \gamma$ lead to different expressions in  unitary and in 
general $R_{\xi}$ gauges. Latter, the results in 
Ref.~\cite{Gegelia:2018pjz} confirm  again the gauge invariance 
of $H\rightarrow\gamma \gamma$. On the other hand, the Higgs production in 
two-photon process and one-loop transition form factor
for $H\rightarrow \gamma \gamma^*$ has 
been computed in Ref.~\cite{Watanabe:2013ria}. 
Furthermore, the Higgs production at $e^-\gamma$ collision via 
the process $e^-\gamma\rightarrow e^- H \rightarrow e^- b \bar{b}$ 
has been considered in Ref.~\cite{Watanabe:2014xaa}. 
To the best of our knowledge, 
there are not available one-loop form factors for decay channel 
$H\rightarrow \gamma^* \gamma^*$. 

In this paper, the detailed calculations for one-loop form factors 
for $H\rightarrow \gamma^* \gamma^*$ in $R_{\xi}$ gauge are 
presented, considering all cases of two on-shell, one on-shell 
and two off-shell for final photons. The computations are 
performed within standard model and in arbitrary beyond the 
standard model (BSM) which the charged scalar particles
may exchange in one-loop Feynman diagrams. 
The analytical results for the form factors are expressed 
in terms of Passarino-Veltman functions which are presented
in standard forms of {\tt LoopTools}~\cite{Hahn:1998yk}. 
Analytic formulas for these functions are well-known and 
their numerical evaluations
can be generated by using {\tt LoopTools}. In our present paper, 
analytic results are shown in $R_{\xi=1}$ for 
$H\rightarrow \gamma^* \gamma^*$ and $\gamma \gamma^*$. 
While one-loop form factor formulas 
for  $H\rightarrow \gamma \gamma$  are presented in both 
't Hooft-Veltman and general $R_{\xi}$ gauges. We also verify
the previous calculations in the case of two on-shell photons. 
The $\xi$-independent 
of the result is also discussed. We show the numerical
checks for one-loop form factors $H\rightarrow \gamma \gamma$ 
with varying $\xi=0,1$ and $\xi\rightarrow \infty$.

The layout of the paper is as follows: In section 2,
we present briefly one-loop tensor reduction method. We then present
the evaluations in detail for one-loop form factors of Higgs decay 
into two photons. Analytical results for the form factors 
with two real photons, one virtual photon, two virtual photons
are shown in this section. Conclusions and outlook are devoted 
in section 3. In appendices, Feynman rules and one-loop amplitude 
for the decay channel are discussed. 
\section{Calculations}
In this calculation, we apply the technique for the reduction of 
one-loop tensor integrals developed in Ref.~\cite{Denner:2005nn}. 
In following section, we describe briefly this approach. In general, 
one-loop one-, 
two- and three-point tensor integrals with rank $P$ are defined as:
\begin{eqnarray}
 \{A; B; C\}^{\mu_1\mu_2\cdots \mu_P}= \int \frac{d^dk}{(2\pi)^d} 
 \dfrac{k^{\mu_1}k^{\mu_2}\cdots k^{\mu_P}}{\{D_1; D_1 D_2; D_1D_2D_3\}}.
\end{eqnarray}
In this formula, 
$D_j$ for $j=1,2,3$ are the inverse Feynman propagators 
which are given:
\begin{eqnarray}
 D_j = (k+ q_j)^2 -m_j^2 +i\rho.
\end{eqnarray}
Where $q_j$ are defined as $q_j = \sum\limits_{i=1}^j p_i$, 
$p_i$ are external momenta; $m_j$ are internal masses. 
The reduction formulas for one-loop one-, two-, three-points 
tensor integrals up to rank $P=3$ are written explicitly 
as follows \cite{Denner:2005nn}:
\begin{eqnarray}
A^{\mu}        &=& 0, \\
A^{\mu\nu}     &=& g^{\mu\nu} \mathbf{A}_{00}, \\
A^{\mu\nu\rho} &=& 0,\\
B^{\mu}        &=& q^{\mu} \mathbf{B}_1,\\
B^{\mu\nu}     &=& g^{\mu\nu} \mathbf{B}_{00} + q^{\mu}q^{\nu} \mathbf{B}_{11}, \\
B^{\mu\nu\rho} &=& \{g, q\}^{\mu\nu\rho} \mathbf{B}_{001} 
+ q^{\mu}q^{\nu}q^{\rho} \mathbf{B}_{111}
\end{eqnarray}
and
\begin{eqnarray}
C^{\mu}        &=& q_1^{\mu} \mathbf{C}_1 + q_2^{\mu} \mathbf{C}_2 
 = \sum\limits_{i=1,2}q_i^{\mu} \mathbf{C}_i \\
 C^{\mu\nu}    &=& g^{\mu\nu} \mathbf{C}_{00} 
 + \sum\limits_{i,j=1,2}q_i^{\mu}q_j^{\nu} \mathbf{C}_{ij}\\
C^{\mu\nu\rho} &=&
	\sum_{i=1}^2 \{g,q_i\}^{\mu\nu\rho} \mathbf{C}_{00i}+
	\sum_{i,j,k=1}^2 q^{\mu}_i q^{\nu}_j q^{\rho}_k \mathbf{C}_{ijk}.
\end{eqnarray}
Here we use the short notation
$\{g, q_i\}^{\mu\nu\rho} = g^{\mu\nu} q^{\rho}_i 
+ g^{\nu\rho} q^{\mu}_i + g^{\mu\rho} q^{\nu}_i$. 
In this method, scalar coefficients 
$\mathbf{A}_{00}, \mathbf{B}_1, \cdots, \mathbf{C}_{222}$
in right hand side of the above relations are so-called 
Passarino-Veltman functions~\cite{Denner:2005nn, Hahn:1998yk}. 
The analytical results for these functions are well-known and 
implemented into computer program named 
{\tt LoopTools}~\cite{Hahn:1998yk} for numerical evaluations.

\begin{center}
\begin{figure}[h]
\includegraphics[scale=1.0]{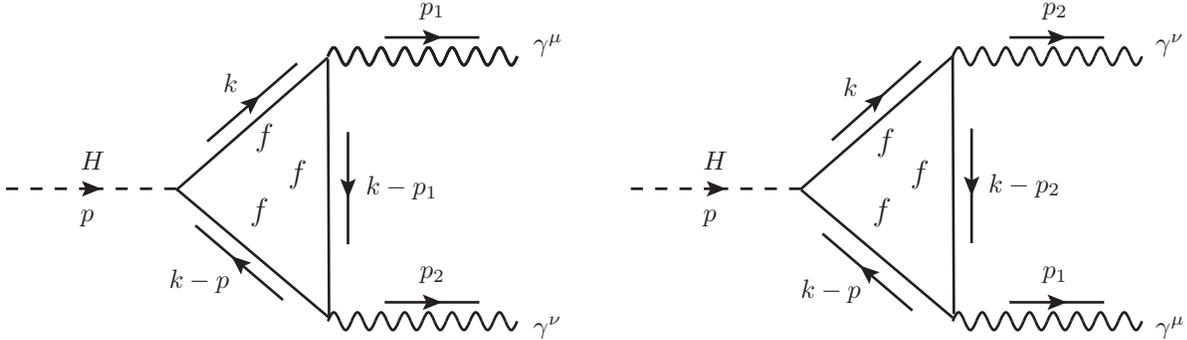}
\caption{\label{fem} Fermion loop Feynman diagrams 
of $H\rightarrow \gamma\gamma$ in 
$R_{\xi}$ gauge.}
\end{figure}
\end{center}

\begin{center}
\begin{figure}[h]
\includegraphics[scale=1.0]{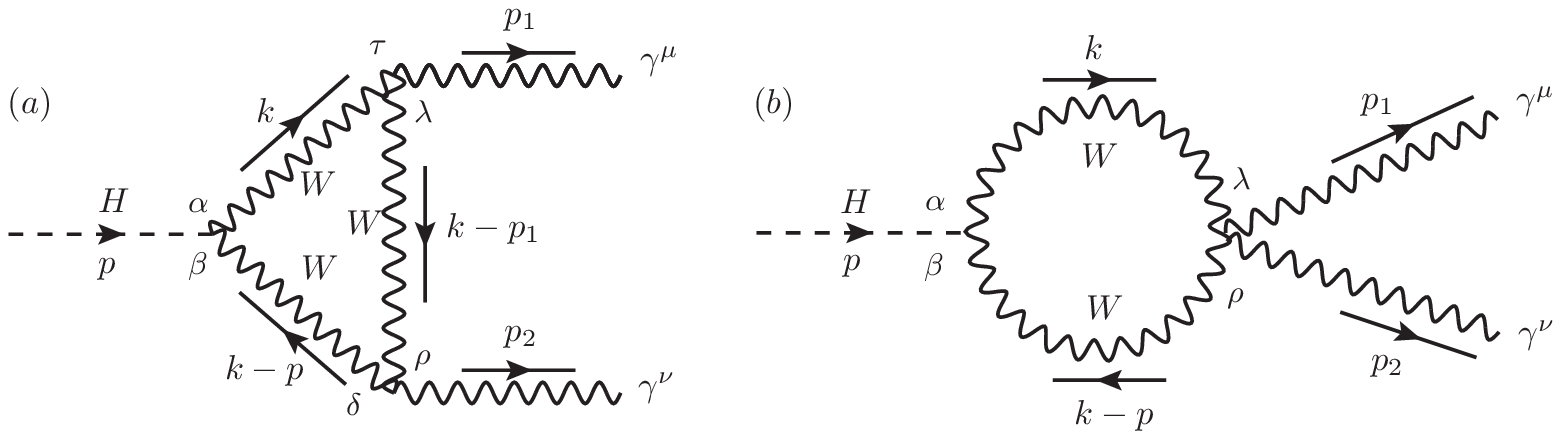}
\includegraphics[scale=1.0]{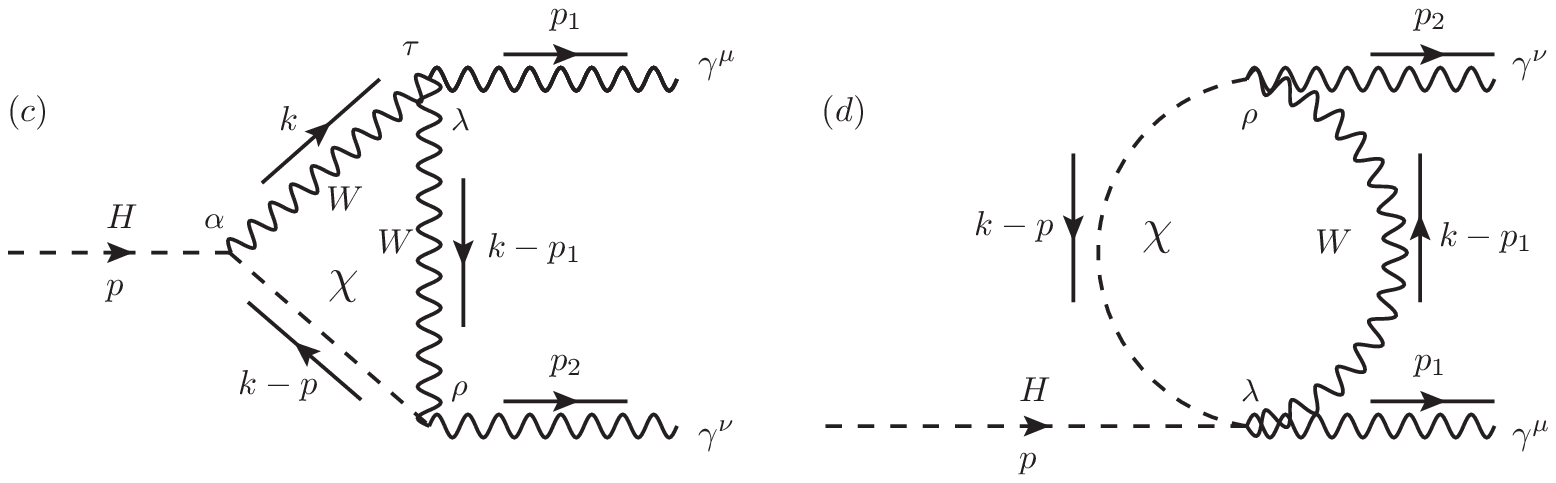}
\includegraphics[scale=1.0]{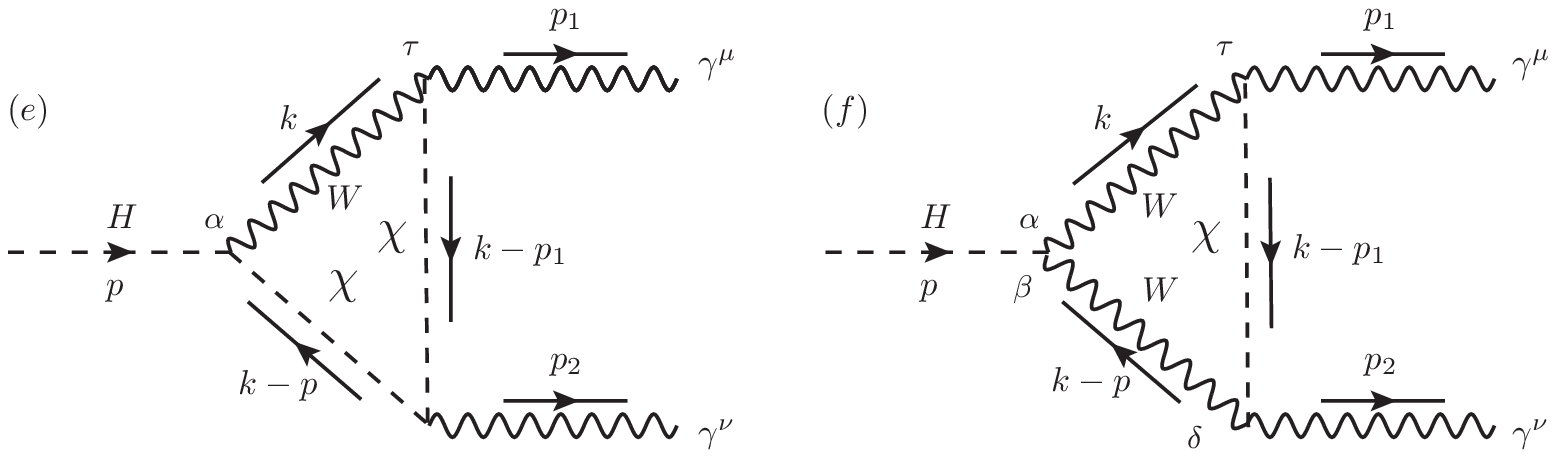}
\includegraphics[scale=1.0]{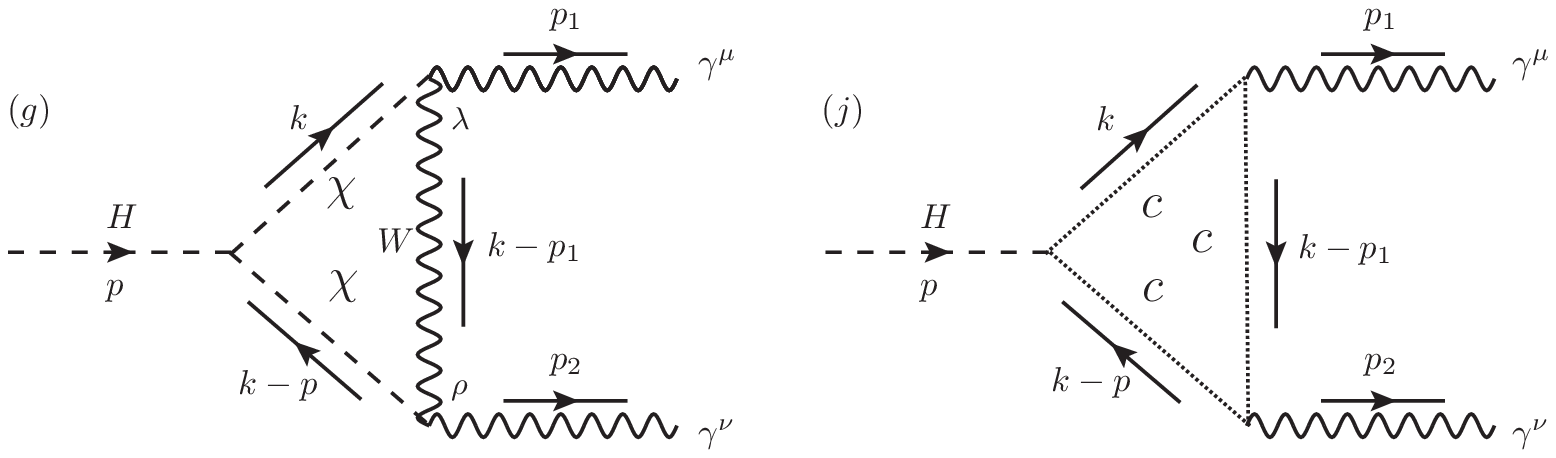}
\includegraphics[scale=1.0]{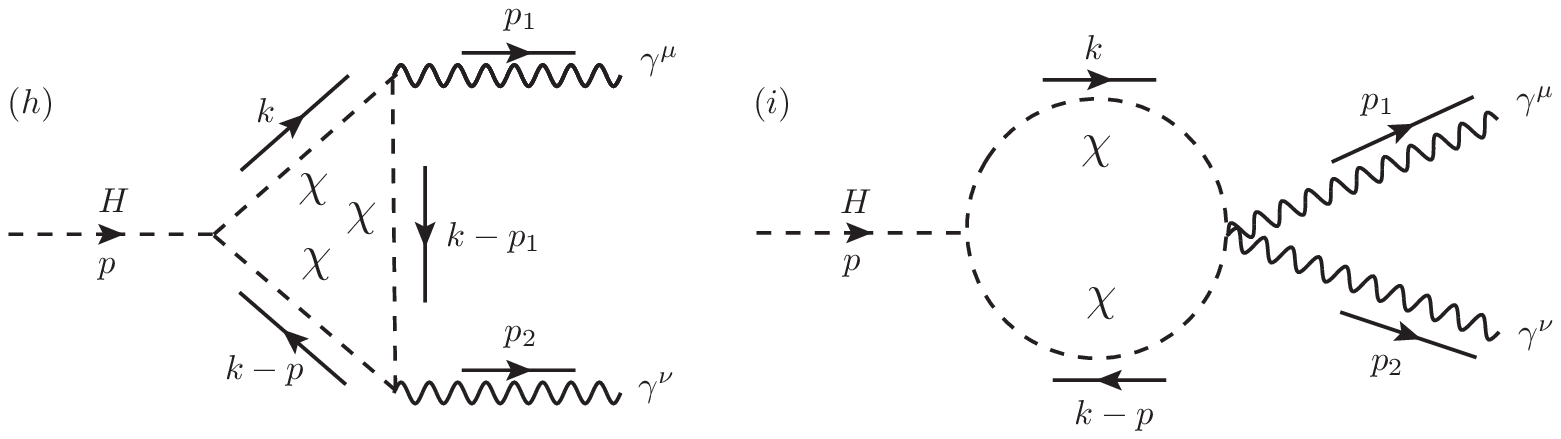}
\caption{\label{wboson} $W$ boson, Goldstone boson, 
Ghost particles exchanging in the loop diagrams  
of $H\rightarrow \gamma \gamma$ in 
$R_{\xi}$ gauge.}
\end{figure}
\end{center}

\begin{center}
\begin{figure}[h]
\includegraphics[scale=1.0]{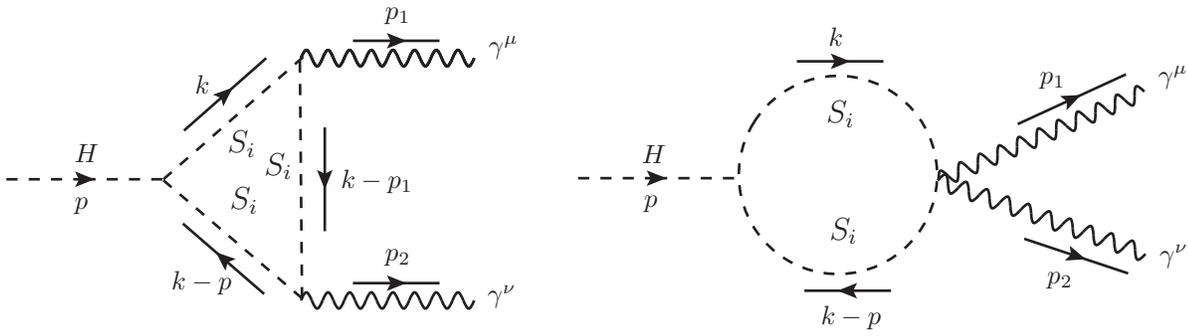}
\caption{\label{charscal} Charged scalar $S_i$ exchanged in 
one-loop Feynman diagrams 
of $H\rightarrow \gamma\gamma$ in 
$R_{\xi}$ gauge.}
\end{figure}
\end{center}

We turn our attention to apply the above approach for
evaluating the decay process $H\rightarrow \gamma^* \gamma^*$. 
Within standard model, the decay channel 
in $R_{\xi}$ consists of fermion loop diagrams
(as shown in Fig.~\ref{fem}) and $W$ boson, Goldstone boson, 
Ghost particles exchanging in the loop diagrams 
(seen Fig.~\ref{wboson}). 
In arbitrary beyond the standard model, we also consider 
the charged scalar particles in the one-loop diagrams 
(described in Fig.~\ref{charscal}).  
In general, the total amplitude of 
the decay $H\rightarrow \gamma^* \gamma^*$ is presented in terms of 
the Lorentz invariant structure as follows:
\begin{eqnarray}
 \mathcal{A}_{H\rightarrow \gamma^* \gamma^*} = \frac{e^2 g}{16\pi^2 M_W} 
 \Big(\mathcal{A}_{00} g^{\mu\nu} + \sum\limits_{i,j=1}^2 \mathcal{A}_{ij}
 p_i^{\mu}p_j^{\nu}  
 \Big) \epsilon_\mu^*(p_1)
 \epsilon_\nu^*(p_2). 
\end{eqnarray}
The kinematic invariant variables involving the decay channel are 
\begin{eqnarray}
 p^2 = (p_1+p_2)^2= M_H^2, \quad p_1^2 \quad \text{and}\quad p_2^2. 
\end{eqnarray}
In this paper, the form factors $\mathcal{A}_{00}, \mathcal{A}_{ij}$
for $i,j=1,2$ are expressed in terms of Passarino-Veltman functions 
mentioned in the beginning of this section.

In general $R_{\xi}$ gauge, in order to simplify
the calculations, $W$ boson propagator is decomposed into 
the following form with a short notation $M_{\xi}^2 = \xi M_W^2$,
\begin{eqnarray}
\label{WProb}
 \dfrac{- i}{p^2 - M_W^2} 
\Bigg[ g^{\mu \nu} - (1 - \xi) 
\dfrac{p^\mu p^\nu}{p^2 - M_{\xi}^2}  \Bigg]
= \dfrac{- i}{p^2 - M_W^2} \left( g^{\mu\nu} 
- \dfrac{k^{\mu}k^{\nu}}{M_W^2} \right)
+ \dfrac{-i}{p^2 - M_{\xi}^2}\dfrac{k^{\mu}k^{\nu}}{M_W^2}.
\end{eqnarray}
The first term in the right hand side of this equation is nothing
but it is $W$ boson propagator in  unitary gauge. While the second
term  relates to Goldstone boson and Ghost particles.  

The calculations 
are performed with the help of {\tt Package-X}~\cite{Patel:2015tea}
for handling all Dirac traces in $d$ dimensions. 
The one-loop form factors are then written in terms of 
Passarino-Veltman functions in standard notations of 
{\tt LoopTools}~\cite{Hahn:1998yk} 
on a diagram-by-diagram basis. 

\subsection{Two off-shell photons}       
We first present analytic results for 
one-loop form factors for the decay 
$H\rightarrow \gamma^* \gamma^*$. The notation
$\gamma^*$ is to one off-shell (or a virtual photon). 
We arrive at the contribution of fermion loop diagrams.
Analytic formulas for the form factors are written
in terms of Passarino-Veltman functions as 
\begin{eqnarray}
 \mathcal{A}_{00}^{(f)}  &=&
-4 m_f^2 N_C Q_f^2  \Big\{
\mathbf{B}_0\left(M_H^2;m_f^2,m_f^2\right)
-4 \mathbf{C}_{00}\left(M_H^2,p_1^2,p_2^2;m_f^2,m_f^2,m_f^2\right)
\nonumber\\
&& +\dfrac{M_H^2-p_1^2-p_2^2}{2}
\; \mathbf{C}_0\left(M_H^2,p_1^2,p_2^2;m_f^2,m_f^2,m_f^2\right)
\Big\},
\\
&& \nonumber\\
\mathcal{A}_{11}^{(f)}  &=& 
8 m_f^2 N_C Q_f^2 
\Big\{
2 \mathbf{C}_{11}\left(M_H^2,p_1^2,p_2^2;m_f^2,m_f^2,m_f^2\right)
+\mathbf{C}_1\left(M_H^2,p_1^2,p_2^2;m_f^2,m_f^2,m_f^2\right)
\Big\},
\\
&& \nonumber\\
\mathcal{A}_{12}^{(f)}  &=& 
4 m_f^2 N_C Q_f^2 
\Big\{ 
4\mathbf{C}_{12}\left(M_H^2,p_1^2,p_2^2;m_f^2,m_f^2,m_f^2\right)
+
4 \mathbf{C}_{11}\left(M_H^2,p_1^2,p_2^2;m_f^2,m_f^2,m_f^2\right)
\nonumber\\
&&
+4 \mathbf{C}_1\left(M_H^2,p_1^2,p_2^2;m_f^2,m_f^2,m_f^2\right)
+2 \mathbf{C}_2\left(M_H^2,p_1^2,p_2^2;m_f^2,m_f^2,m_f^2\right)
\nonumber\\
&&
+\mathbf{C}_0\left(M_H^2,p_1^2,p_2^2;m_f^2,m_f^2,m_f^2\right)
\Big\},
\\
\mathcal{A}_{21}^{(f)}  &=& 
4 m_f^2 N_C Q_f^2 
\Big\{
4\mathbf{C}_{12}\left(M_H^2,p_1^2,p_2^2;m_f^2,m_f^2,m_f^2\right)
+4\mathbf{C}_{11}\left(M_H^2,p_1^2,p_2^2;m_f^2,m_f^2,m_f^2\right)
\nonumber\\
&&
+4\mathbf{C}_1\left(M_H^2,p_1^2,p_2^2;m_f^2,m_f^2,m_f^2\right)
+\mathbf{C}_0\left(M_H^2,p_1^2,p_2^2;m_f^2,m_f^2,m_f^2\right)
\Big\},
\\
\mathcal{A}_{22}^{(f)}  &=& 
8 m_f^2 N_C Q_f^2 
\Big\{
2 \mathbf{C}_{22}\left(M_H^2,p_1^2,p_2^2;m_f^2,m_f^2,m_f^2\right)
+4\mathbf{C}_{12}\left(M_H^2,p_1^2,p_2^2;m_f^2,m_f^2,m_f^2\right)
\nonumber\\
&&
+2 \mathbf{C}_{11}\left(M_H^2,p_1^2,p_2^2;m_f^2,m_f^2,m_f^2\right)
+\mathbf{C}_0\left(M_H^2,p_1^2,p_2^2;m_f^2,m_f^2,m_f^2\right)
\nonumber\\
&&
+3 \mathbf{C}_2\left(M_H^2,p_1^2,p_2^2;m_f^2,m_f^2,m_f^2\right)
+3 \mathbf{C}_1\left(M_H^2,p_1^2,p_2^2;m_f^2,m_f^2,m_f^2\right)
\Big\}.
\end{eqnarray}
Here $N_C$ is a color factor ($N_C=1$ for leptons and
$N_C=3$ for quarks) and $Q_fe$ is electric charge of 
fermions.  

We next consider $W$ boson contributions for 
the form factors. In general $R_{\xi}$ gauge, 
the contributions are included $W$ boson, Goldstone 
boson and Ghost particles in one-loop diagrams. 
Summing all these diagrams, we get the form factors
which are functions of the unphysical parameter $\xi$ and the
kinematic invariants
$p_1^2,p_2^2,M_H^2,M_W^2$. For illustrating, we only 
show here the results in 't Hooft-Veltman gauge:
\begin{eqnarray}
 \mathcal{A}_{00}^{(W)} &=&
 M_W^2 
 \Big\{
 -\mathbf{B}_0\left(p_1^2;M_W^2,M_W^2\right)
 +\mathbf{B}_0\left(p_2^2;M_W^2,M_W^2\right)
\n \\
 &&
 +2 \left(-4 M_H^2+p_1^2+4 p_2^2\right) 
 \mathbf{C}_0\left(p_1^2,p_2^2,M_H^2;M_W^2,M_W^2,M_W^2\right)
 \Big\}
 \nonumber\\
&& 
 -\left[M_H^2+2 (d-1) M_W^2\right] \times
 \\
 &&\times
 \Big\{
 \mathbf{B}_0\left(M_H^2;M_W^2,M_W^2\right)
 -4\mathbf{C}_{00}\left(p_1^2,p_2^2,M_H^2;M_W^2,M_W^2,M_W^2\right)
 \Big\},
 \nonumber\\
 \mathcal{A}_{11}^{(W)} &=&
 2\left[M_H^2+2 (d+2) M_W^2\right] 
 \mathbf{C}_0\left(p_1^2,p_2^2,M_H^2;M_W^2,M_W^2,M_W^2\right)
 \nonumber\\
 &&
 +2\left[3 M_H^2+(6 d-1) M_W^2\right]
 \Big\{
 \mathbf{C}_2\left(p_1^2,p_2^2,M_H^2;M_W^2,M_W^2,M_W^2\right)
 \nonumber\\
 &&
 +\mathbf{C}_1\left(p_1^2,p_2^2,M_H^2;M_W^2,M_W^2,M_W^2\right)
 \Big\}
 \nonumber\\
 &&
 + 4 \left[M_H^2+2 (d-1) M_W^2\right]
 \Big\{
 \mathbf{C}_{11}\left(p_1^2,p_2^2,M_H^2;M_W^2,M_W^2,M_W^2\right)
 \nonumber\\
 &&
 +
 \mathbf{C}_{22}\left(p_1^2,p_2^2,M_H^2;M_W^2,M_W^2,M_W^2\right)
 +2 \mathbf{C}_{12}\left(p_1^2,p_2^2,M_H^2;M_W^2,M_W^2,M_W^2\right)
 \Big\},
 \\
 \mathcal{A}_{12}^{(W)} &=&
 4 \left[M_H^2+2 (d-1) M_W^2\right]\times
 \nonumber\\
 &&
 \times
 \Big[
 \mathbf{C}_{22}\left(p_1^2,p_2^2,M_H^2;M_W^2,M_W^2,M_W^2\right)
 +
 \mathbf{C}_{12}\left(p_1^2,p_2^2,M_H^2;M_W^2,M_W^2,M_W^2\right)
 \Big]
 \nonumber\\
 &&
 +2 \left[M_H^2+2 d M_W^2\right] 
 \nonumber\\
 &&
 \times
 \Big[
 2 \mathbf{C}_2\left(p_1^2,p_2^2,M_H^2;M_W^2,M_W^2,M_W^2\right)
 +
 \mathbf{C}_1\left(p_1^2,p_2^2,M_H^2;M_W^2,M_W^2,M_W^2\right)
 \Big]
 \nonumber\\
 &&
 +\left[M_H^2+2 (d+1) M_W^2\right]
 \mathbf{C}_0\left(p_1^2,p_2^2,M_H^2;M_W^2,M_W^2,M_W^2\right),
 \\
 \mathcal{A}_{21}^{(W)} &=&
2 M_W^2 
\Big\{ 4 (d-1) 
\Big[
\mathbf{C}_{22}\left(p_1^2,p_2^2,M_H^2;M_W^2,M_W^2,M_W^2\right)
\nonumber\\
&&
+\mathbf{C}_{12}\left(p_1^2,p_2^2,M_H^2;M_W^2,M_W^2,M_W^2\right)
\Big]
\nonumber\\
&&
+(4 d+2) \mathbf{C}_2\left(p_1^2,p_2^2,M_H^2;M_W^2,M_W^2,M_W^2\right)
+11 \mathbf{C}_0\left(p_1^2,p_2^2,M_H^2;M_W^2,M_W^2,M_W^2\right)
\nonumber\\
&&
+3 \mathbf{C}_1\left(p_1^2,p_2^2,M_H^2;M_W^2,M_W^2,M_W^2\right)
\Big\}
+4 M_H^2 
\Big\{
\mathbf{C}_{22}\left(p_1^2,p_2^2,M_H^2;M_W^2,M_W^2,M_W^2\right)
\nonumber\\
&&
+\mathbf{C}_{12}\left(p_1^2,p_2^2,M_H^2;M_W^2,M_W^2,M_W^2\right)+\mathbf{C}_2\left(p_1^2,p_2^2,M_H^2;M_W^2,M_W^2,M_W^2\right)
\Big\},
\\
\mathcal{A}_{22}^{(W)}&=&
4 \left[M_H^2+2 (d-1) M_W^2\right] 
\mathbf{C}_{22}\left(p_1^2,p_2^2,M_H^2;M_W^2,M_W^2,M_W^2\right)
\nonumber\\
&&
+2 \left[M_H^2+(2 d+3) M_W^2\right] 
\mathbf{C}_2\left(p_1^2,p_2^2,M_H^2;M_W^2,M_W^2,M_W^2\right)
\nonumber\\
&&
+4 M_W^2 \mathbf{C}_0\left(p_1^2,p_2^2,M_H^2;M_W^2,M_W^2,M_W^2\right).
\end{eqnarray}

We extend our calculation with considering the charged 
scalar particles
$S_i$ exchanged in one-loop diagrams.
The amplitude for $H\rightarrow \gamma^*\gamma^*$ 
with exchanging the charged scalar particles 
in loop diagrams are decomposed
\begin{eqnarray}
 \mathcal{A}_{H\rightarrow \gamma\gamma}^{(S_i)} = \frac{\lambda_{HS_iS_i} e^2Q_{S_i}^2}{16\pi^2 } 
 \Big(\mathcal{A}_{00}^{(S_i)} g^{\mu\nu} + \sum\limits_{i,j=1}^2 \mathcal{A}_{ij}^{(S_i)}
 p_i^{\mu}p_j^{\nu}  
 \Big) \epsilon_\mu^*(p_1)
 \epsilon_\nu^*(p_2). 
\end{eqnarray}
Applying the same procedure, we derive 
one-loop form factors due to the contributions of the charged scalar 
loop diagrams as follows:

\begin{eqnarray}
\mathcal{A}^{(S_i)}_{00} &=&
-2 \Big\{
\mathbf{B}_0\left(M_H^2;M_{S_i}^2,M_{S_i}^2\right)-4
\mathbf{C}_{00}\left(M_H^2,p_1^2,p_2^2;M_{S_i}^2,M_{S_i}^2,M_{S_i}^2\right)
\Big\}, \\
\mathcal{A}^{(S_i)}_{11} &=&
4\Big\{ 
2 \mathbf{C}_{11}\left(M_H^2,p_1^2,p_2^2;M_{S_i}^2,M_{S_i}^2,M_{S_i}^2\right)
+\mathbf{C}_1\left(M_H^2,p_1^2,p_2^2;M_{S_i}^2,M_{S_i}^2,M_{S_i}^2\right)
\Big\},   
\\
\mathcal{A}^{(S_i)}_{12} &=& 2\Big\{
4\mathbf{C}_{12}\left(M_H^2,p_1^2,p_2^2;M_{S_i}^2,M_{S_i}^2,M_{S_i}^2\right)
+4\mathbf{C}_{11}\left(M_H^2,p_1^2,p_2^2;M_{S_i}^2,M_{S_i}^2,M_{S_i}^2\right)
\nonumber\\
&&
+4\mathbf{C}_1\left(M_H^2,p_1^2,p_2^2;M_{S_i}^2,M_{S_i}^2,M_{S_i}^2\right)
+2 \mathbf{C}_2\left(M_H^2,p_1^2,p_2^2;M_{S_i}^2,M_{S_i}^2,M_{S_i}^2\right)
\nonumber\\
&&
+\mathbf{C}_0\left(M_H^2,p_1^2,p_2^2;M_{S_i}^2,M_{S_i}^2,M_{S_i}^2\right)
\Big\}
\end{eqnarray}
and 
\begin{eqnarray}
\mathcal{A}^{(S_i)}_{21} &=&
8\Big\{
\mathbf{C}_{12}\left(M_H^2,p_1^2,p_2^2;M_{S_i}^2,M_{S_i}^2,M_{S_i}^2\right)
+\mathbf{C}_{11}\left(M_H^2,p_1^2,p_2^2;M_{S_i}^2,M_{S_i}^2,M_{S_i}^2\right)
\nonumber\\
&&
+\mathbf{C}_1\left(M_H^2,p_1^2,p_2^2;M_{S_i}^2,M_{S_i}^2,M_{S_i}^2\right)
\Big\},
\\
\mathcal{A}^{(S_i)}_{22} &=&
4\Big\{
2 \mathbf{C}_{22}\left(M_H^2,p_1^2,p_2^2;M_{S_i}^2,M_{S_i}^2,M_{S_i}^2\right)
+4 \mathbf{C}_{12}\left(M_H^2,p_1^2,p_2^2;M_{S_i}^2,M_{S_i}^2,M_{S_i}^2\right)
\nonumber\\
&&
+2 \mathbf{C}_{11}\left(M_H^2,p_1^2,p_2^2;M_{S_i}^2,M_{S_i}^2,M_{S_i}^2\right)
+\mathbf{C}_0\left(M_H^2,p_1^2,p_2^2;M_{S_i}^2,M_{S_i}^2,M_{S_i}^2\right)
\nonumber\\
&&
+3 \mathbf{C}_2\left(M_H^2,p_1^2,p_2^2;M_{S_i}^2,M_{S_i}^2,M_{S_i}^2\right)
+3 \mathbf{C}_1\left(M_H^2,p_1^2,p_2^2;M_{S_i}^2,M_{S_i}^2,M_{S_i}^2\right)
\Big\}.   
\end{eqnarray}
By taking the limit of $p_1^2\rightarrow 0$ or $p_2^2\rightarrow 0$, we 
get the results for the case of one off-shell photon. We refer analytic results
for all form factors in which $\gamma(p_1)$ is on-shell photon in appendix 
$A$. 
\subsection{Two on-shell photons}       
We change our topic to the case of two real 
photons in the final state of this channel.
In this case, on-shell conditions for these photons 
are implied as 
\begin{eqnarray}
 p_1^2=p_2^2=0. 
\end{eqnarray}
We then have the relation $p_1p_2 =M_H^2/2$. 
On the other hand, Ward identities are taken 
into account for both photons
\begin{eqnarray}
p_1^{\mu}\epsilon^*_{\mu}(p_1)=
p_2^{\nu}\epsilon^*_{\nu}(p_2)=0. 
\end{eqnarray}
Subsequently, all form factors 
\begin{eqnarray}
 \mathcal{A}_{11}=\mathcal{A}_{12}=\mathcal{A}_{22}=0.
\end{eqnarray}
Analytic formulas for the remaining
form factors are derived as: 
\begin{eqnarray}
 \mathcal{A}_{00}^{(f)}  &=&
-2 m_f^2 N_C Q_f^2 
\Big\{
2 \mathbf{B}_0\left(M_H^2;m_f^2,m_f^2\right)-8
\mathbf{C}_{00}\left(M_H^2,0,0;m_f^2,m_f^2,m_f^2\right)
\nonumber\\
&&
+M_H^2 \mathbf{C}_0\left(M_H^2,0,0;m_f^2,m_f^2,m_f^2\right)
\Big\},
\\ 
&& \nonumber\\
\mathcal{A}_{21}^{(f)}  &=& 
4 m_f^2 N_C Q_f^2 \Big\{
4\Big[
\mathbf{C}_{12}\left(M_H^2,0,0;m_f^2,m_f^2,m_f^2\right)
+\mathbf{C}_{11}\left(M_H^2,0,0;m_f^2,m_f^2,m_f^2\right)
\nonumber\\
&&
+\mathbf{C}_1\left(M_H^2,0,0;m_f^2,m_f^2,m_f^2\right)
\Big]
+\mathbf{C}_0\left(M_H^2,0,0;m_f^2,m_f^2,m_f^2\right)
\Big\}.
\end{eqnarray}
From one-loop $W$ boson contributions, the analytical results
in both 't Hooft-Veltman and general $R_{\xi}$ gauges 
are shown. First, the form factors in $R_{\xi}$ gauge 
read
\begin{eqnarray}
\mathcal{A}^{(W)}_{00}(\xi)&=&
\Big[
2 M_W^2 (1-d) -2 M_H^2 - \dfrac{M_H^4}{2 M_W^2} 
\Big] 
\mathbf{B}_0(M_H^2;M_W^2,M_W^2)
\n \\
&&
+\Big[
\dfrac{M_H^4}{M_W^2}+M_H^2(1 - \xi)
\Big]
 \mathbf{B}_0(M_H^2;M_{\xi}^2,M_W^2)
 \n \\
&&
-\Big(\dfrac{M_H^4}{M_W^2}+2 M_H^2 \Big) \mathbf{B}_1(M_H^2;M_W^2,M_W^2)
\n \\
&&
+ \Big(M_H^2 \xi-\dfrac{M_H^4}{2 M_W^2} \Big) 
\Big[2 \mathbf{B}_1 + \mathbf{B}_0 \Big] (M_H^2;M_{\xi}^2,M_{\xi}^2)
 \\
&&
+\Big[\dfrac{M_H^4}{M_W^2} + M_H^2 (1-\xi) \Big] \Big[ \mathbf{B}_1(M_H^2;M_W^2,M_{\xi}^2) + \mathbf{B}_1(M_H^2;M_{\xi}^2,M_W^2) \Big]
\n \\
&&
+[4 M_H^2+8 M_W^2 (d-1)] \mathbf{C}_{00}(0,0,M_H^2;M_W^2,M_W^2,M_W^2)
\n \\
&&
+\Big[ M_H^2 +M_W^2 (1-\xi) \Big] \times
\n \\
&&
\times
\Big[
2\mathbf{C}_{00}(0,0,M_H^2;M_W^2,M_W^2,M_{\xi}^2) 
-2 \mathbf{C}_{00}(0,0,M_H^2;M_{\xi}^2,M_W^2,M_W^2)
\n \\
&&
+\mathbf{C}_{00}(0,0,M_H^2;M_W^2,M_{\xi}^2,M_{\xi}^2)
-\mathbf{C}_{00}(0,0,M_H^2;M_{\xi}^2,M_{\xi}^2,M_W^2)
\Big]
\n \\
&&
+2 M_H^2 M_W^2 
\Big[
\mathbf{C}_0(0,0,M_H^2;M_{\xi}^2,M_W^2,M_W^2)
\n \\
&&
-\mathbf{C}_0(0,0,M_H^2;M_W^2,M_W^2,M_{\xi}^2)
-4 \mathbf{C}_0(0,0,M_H^2;M_W^2,M_W^2,M_W^2)
\Big],
\n \\
&&\n \\
\mathcal{A}^{(W)}_{21}(\xi)&=&
(4 M_H^2+8 d M_W^2-8 M_W^2) 
\Big[ 
\mathbf{C}_{22} 
+\mathbf{C}_{12}
+\mathbf{C}_2 
\Big] (0,0,M_H^2;M_W^2,M_W^2,M_W^2)
\n \\
&&
+\Big[M_H^2 + M_W^2 (1-\xi)\Big] \times
\n \\
&&
\times
\Big[
2\mathbf{C}_{22}(0,0,M_H^2;M_W^2,M_W^2,M_{\xi}^2)
-2\mathbf{C}_{22}(0,0,M_H^2;M_{\xi}^2,M_W^2,M_W^2)
\n \\
&&
+2\mathbf{C}_{12}(0,0,M_H^2;M_W^2,M_W^2,M_{\xi}^2)
-2\mathbf{C}_{12}(0,0,M_H^2;M_{\xi}^2,M_W^2,M_W^2)
\n \\
&&
+\mathbf{C}_{22}(0,0,M_H^2;M_W^2,M_{\xi}^2,M_{\xi}^2)
-\mathbf{C}_{22}(0,0,M_H^2;M_{\xi}^2,M_{\xi}^2,M_W^2)
\n \\
&&
+\mathbf{C}_{12}(0,0,M_H^2;M_W^2,M_{\xi}^2,M_{\xi}^2)
-\mathbf{C}_{12}(0,0,M_H^2;M_{\xi}^2,M_{\xi}^2,M_W^2)
\Big]
\n \\
&&
+2 M_W^2 
\Big[
2\mathbf{C}_1(0,0,M_H^2;M_{\xi}^2,M_W^2,M_W^2)
+\mathbf{C}_1(0,0,M_H^2;M_W^2,M_{\xi}^2,M_{\xi}^2)
\n \\
&&
+2\mathbf{C}_0(0,0,M_H^2;M_W^2,M_W^2,M_{\xi}^2)
+\mathbf{C}_0(0,0,M_H^2;M_W^2,M_{\xi}^2,M_{\xi}^2)
\Big]
\n \\
&&
+\Big[M_H^2 + M_W^2 (3-\xi)\Big] \times
\\
&&
\times
\Big[
2\mathbf{C}_2(0,0,M_H^2;M_W^2,M_W^2,M_{\xi}^2)
+\mathbf{C}_2(0,0,M_H^2;M_W^2,M_{\xi}^2,M_{\xi}^2)
\Big]
\n \\
&&
+\Big[M_W^2 (1+\xi) - M_H^2 \Big] \times
\n \\
&&
\times
\Big[
2\mathbf{C}_2(0,0,M_H^2;M_{\xi}^2,M_W^2,M_W^2)
+\mathbf{C}_2(0,0,M_H^2;M_{\xi}^2,M_{\xi}^2,M_W^2)
\Big]
\n \\
&&
+16 M_W^2 \mathbf{C}_0(0,0,M_H^2;M_W^2,M_W^2,M_W^2)
\n
\end{eqnarray}
By setting $\xi=1$, we obtain the results  
in 't Hooft-Veltman gauge
\begin{eqnarray}
\mathcal{A}_{00}^{(W)}&=&
-\left[M_H^2+2 (d-1) M_W^2\right]
\Big[
\mathbf{B}_0\left(M_H^2;M_W^2,M_W^2\right)
-4 \mathbf{C}_{00}\left(0,0,M_H^2;M_W^2,M_W^2,M_W^2\right)
\Big]
\nonumber\\
&&
-8 M_H^2 M_W^2 \mathbf{C}_0\left(0,0,M_H^2;M_W^2,M_W^2,M_W^2\right),
\\
\mathcal{A}_{21}^{(W)}&=&
2 M_W^2 
\Big\{ 4 (d-1) \Big[ 
\mathbf{C}_{22}\left(0,0,M_H^2;M_W^2,M_W^2,M_W^2\right)
+
\mathbf{C}_{12}\left(0,0,M_H^2;M_W^2,M_W^2,M_W^2\right) \Big]
\nonumber\\
&&
+2(2d+1) \mathbf{C}_2\left(0,0,M_H^2;M_W^2,M_W^2,M_W^2\right)
+11 \mathbf{C}_0\left(0,0,M_H^2;M_W^2,M_W^2,M_W^2\right)
\nonumber\\
&&
+3 \mathbf{C}_1\left(0,0,M_H^2;M_W^2,M_W^2,M_W^2\right) \Big\}
+4 M_H^2 \Big[ 
\mathbf{C}_{22}\left(0,0,M_H^2;M_W^2,M_W^2,M_W^2\right)
\nonumber\\
&&
+\mathbf{C}_{12}\left(0,0,M_H^2;M_W^2,M_W^2,M_W^2\right)
+\mathbf{C}_2\left(0,0,M_H^2;M_W^2,M_W^2,M_W^2\right)\Big].
\end{eqnarray}
One-loop form factors for this process with including
charged scalars in the loop diagrams are shown
\begin{eqnarray}
\mathcal{A}^{(S_i)}_{00} &=&
-2 \Big\{
\mathbf{B}_0\left(M_H^2;M_{S_i}^2,M_{S_i}^2\right)-4
\mathbf{C}_{00}\left(M_H^2,0,0;M_{S_i}^2,M_{S_i}^2,M_{S_i}^2\right)
\Big\}, \\
\mathcal{A}^{(S_i)}_{21} &=&
8\Big\{
\mathbf{C}_{12}\left(M_H^2,0,0;M_{S_i}^2,M_{S_i}^2,M_{S_i}^2\right)
+\mathbf{C}_{11}\left(M_H^2,0,0;M_{S_i}^2,M_{S_i}^2,M_{S_i}^2\right)
\nonumber\\
&&
+\mathbf{C}_1\left(M_H^2,0,0;M_{S_i}^2,M_{S_i}^2,M_{S_i}^2\right)
\Big\}.   
\end{eqnarray}
We find that all form factors in $R_{\xi=1}$ in two on-shell photons
case can be obtained  by taking $p_1^2, p_2^2\rightarrow0$ 
from the results in previous subsection. 

In the limit of 
$d\rightarrow 4$, we confirm previous results, taking
Ref.~\cite{Marciano:2011gm} as an example. 
In detail, our results when $d\rightarrow 4$ are presented
\begin{eqnarray}
\mathcal{A}^{(f)}_{00} &=&  
\mathcal{A}^{(f)}_{21} 
\times\left(-\dfrac{M_H^2}{2}\right) =\\
&=&
\frac{m_f^2 N_CQ_f^2 }{M_H^2}
\left\{
4 M_H^2 +
\left(4 m_f^2-M_H^2\right) \ln^2\left(
\dfrac{-M_H^2+2 m_f^2+\sqrt{M_H^4-4 m_f^2 M_H^2}}{2 m_f^2}\right)
\right\}\nonumber
\end{eqnarray}
and 
\begin{eqnarray}
 \mathcal{A}^{(W)}_{00}
 &=&
 \mathcal{A}^{(W)}_{21}\times\left(-\dfrac{M_H^2}{2}\right)  =\\
 &=&
 M_H^2+6 M_W^2
 +\left(\frac{6 M_W^4}{M_H^2}-3 M_W^2\right) \
 \ln^2\left(\frac{-M_H^2 +2 M_W^2 +\sqrt{M_H^4-4 M_H^2 M_W^2}}{2 M_W^2}\right).
 \nonumber
\end{eqnarray}
These results agree with Ref.~\cite{Marciano:2011gm}.

Furthermore, we also have analytic results 
for the form factors due to the charged scalar in the loop at $d=4$. 
These factors read
\begin{eqnarray}
\mathcal{A}^{(S_i)}_{00}
 &=&
 \mathcal{A}^{(S_i)}_{21}
 \times\left(-\dfrac{M_H^2}{2}\right)  =\\
 &=&
 \frac{2 \lambda_{HS_iS_i} M_W Q_{S_i}^2 
 }{g M_H^2}
 \left\{ 
 M_H^2+M_{S_i}^2 \ln^2\left(\frac{-M_H^2+2 M_{S_i}^2+\sqrt{M_H^4-4 M_H^2 M_{S_i}^2}}{2 M_{S_i}^2}\right)
 \right\}.
 \nonumber
\end{eqnarray}

The $\xi$-independent 
of the result is also discussed. 
The numerical results 
are generated by varying $\xi \rightarrow 0$ (is so-called 
Coulomb gauge), $\xi =1$ or 't Hooft-Veltman gauge 
and $\xi\rightarrow \infty$ ( unitary gauge).  
In this Table \ref{numRE}, we show the numerical results 
of 
\begin{eqnarray}
(-2/M_H^2) \times \mathcal{A}_{00} 
= \mathcal{A}_{21}.
\end{eqnarray}
We find that numerical results are good stability in different gauges.
\begin{table}[h]
\scriptsize
\begin{center}
\begin{tabular}{cllll}  \hline \hline \\
$\textbf{diagrams} / \xi$ & $\xi \rightarrow 0$ & $\xi = 1$ & $\xi =  100$ & $\xi \rightarrow \infty$ \\ \\ \hline \hline
\\
\centering $a$  &  $-3.468876070276491$ & $-4.882018101498933$ &  $-30.40120343875694$ & $-2.777777777516398 
\cdot  10^{11}$ 
\\ & $+ 0.044696724580243 \,i$ &  &  &  \\ \\
\hline
\\
\centering $b$  & $0.4359747855634294$ & $0$ &  $22.87057225068166$ & $2.777777777438721 \cdot10^{11}$ 
\\ & $+ 1.1775870133864408 \,i$ &  &  &  \\ \\
\hline
\\
\centering $c$  &  $-2.288315292217131$ & $-2.345059943153266$ & $-0.5672364405722673$ & $-0.3888888890439217$ 
\\ & $- 0.647907755004331 \,i$ &  &  &  \\ \\
\hline
\\
\centering $d$  &  $0$ & $0$ & $0$ & $0$ 
\\ &  &  &  &  \\
\hline
\\
\centering $e$  &  $-1.599294975435531$ & $-1.591326628583693$ & $-0.1930905308944328$ & $-0.1666666666694787$ 
\\ & $- 1.787032617316791 \,i$ &  &  &  \\ \\
\hline 
\\
\centering $f$  &  $0.2615610107425627$ & $0$ & $0.04694802196574567$ & $0.08333333330075746$ 
\\ & $+ 0.8935163086583953 \,i$ &  &  &  \\ \\
\hline
\\
\centering $g$  &  $0.3357098311660154$ & $0$ & $0.0003307252303052892$ & $3.357186222008115 \cdot 10^{-14}$ 
\\ & $+ 0.3191403256960426 \,i$ &  &  &  \\ \\
\hline
\\
\centering $h$  &  $-2$ & $0.6243029825430905$ & $0.004041657968703369$ & $4.028623466417146 \cdot 10^{-13}$ 
\\ &  &  &  &  \\
\hline
\\
\centering $i$  &  $0$ & $0$ & $0$ & $0$ 
\\ &  &  &  &  \\
\hline
\\
\centering $j$  &  $0$ & $-0.12913901976434381$ & $-0.08360295607991542$ & $-0.08333333333336019$ 
\\ &  &  &  &  \\
\hline
\\
\centering Sum  &  $-8.323240710457146$ & $-8.323240710457146$ 
& $-8.323240710457146$ & $-8.323240710457146$ 
\\ &  &  &  &  \\
\hline\hline
\end{tabular}
\caption{\label{numRE} The numerical checks for the form factors  in the case of $\xi \rightarrow 0, 
\xi =1, \xi\rightarrow \infty$ are shown. For this check, we set  
$M_H = 125$ GeV, $M_W = 80.4$ GeV, and $p_1^2 = p_2^2 = 0$ GeV.  }
\end{center}
\end{table}

\section{Conclusions}   
In this paper, we have presented one-loop form factors 
for $H\rightarrow \gamma^* \gamma^*$ in $R_{\xi}$ gauge, 
considering all cases of two on-shell, one on-shell and 
two off-shell for final photons. The calculations are performed 
in standard model and in arbitrary beyond the standard models  
which the charged scalar particles may be exchanged in 
one-loop diagrams. Analytic results for the form factors are
shown in general forms which are expressed in terms of 
the Passarino-Veltman functions in stadard notation of {\tt LoopTools}. 
We have also confirmed the results in previous computations 
which are available for the case of two on-shell photons. 
The $\xi$-independent of the result has been also studied. 
We find that numerical results are good stability with 
varying $\xi=0,1$ and $\xi\rightarrow \infty$.  
\\

\noindent
{\bf Acknowledgment:}~
This research is funded by Vietnam National Foundation for Science and 
Technology Development (NAFOSTED) under the grant number $103.01$-$2019.346$.\\ 

\section*{Appendix A: One-loop form factors
for $H\rightarrow \gamma\gamma^*$}       
Analytical results for one off-shell photon in the 
decay of $H\rightarrow \gamma\gamma^*$ are reported in this subsection.
Without loss the generality, we consider $\gamma(p_1)$ is real photon.
As a result, we have on-shell condition for $\gamma(p_1)$
which is $p_1^2=0$. Following Ward identity, one also has 
$p_1^{\mu}\epsilon^*_{\mu}(p_1)=0$.  Subsequently, the form factors
$\mathcal{A}_{11}= \mathcal{A}_{12}= 0$. Other form factors
are given in the following paragraphs. Due to the 
fermion loop contributions, the form factors  are shown
\begin{eqnarray} 
 \mathcal{A}_{00}^{(f)}  &=&
 -4 m_f^2 N_C Q_f^2 
 \Big\{
 \mathbf{B}_0\left(M_H^2;m_f^2,m_f^2\right)
 -4 \mathbf{C}_{00}\left(M_H^2,0,p_2^2; m_f^2,m_f^2,m_f^2\right)
 \nonumber\\
 && +\dfrac{M_H^2-p_2^2}{2} \mathbf{C}_0\left(M_H^2,0,p_2^2;m_f^2,m_f^2,m_f^2\right)
\Big\},
\\ 
&& \nonumber\\
\mathcal{A}_{21}^{(f)}  &=& 
4 m_f^2 N_C Q_f^2 
\Big\{
4 \Big[
\mathbf{C}_{12}\left(M_H^2,0,p_2^2;m_f^2,m_f^2,m_f^2\right)
+\mathbf{C}_{11}\left(M_H^2,0,p_2^2;m_f^2,m_f^2,m_f^2\right)
\nonumber\\
&& +\mathbf{C}_1\left(M_H^2,0,p_2^2;m_f^2,m_f^2,m_f^2\right)
\Big]
+\mathbf{C}_0\left(M_H^2,0,p_2^2;m_f^2,m_f^2,m_f^2\right)
\Big\}
\\
&& \nonumber\\   
\mathcal{A}_{22}^{(f)}  &=& 
8 m_f^2 N_C Q_f^2 
\Big\{
2 \mathbf{C}_{22}\left(M_H^2,0,p_2^2;m_f^2,m_f^2,m_f^2\right)
+4 \mathbf{C}_{12}\left(M_H^2,0,p_2^2;m_f^2,m_f^2,m_f^2\right)
\nonumber\\
&&
+2 \mathbf{C}_{11}\left(M_H^2,0,p_2^2;m_f^2,m_f^2,m_f^2\right)
+\mathbf{C}_0\left(M_H^2,0,p_2^2;m_f^2,m_f^2,m_f^2\right)
\nonumber\\
&&
+3 \mathbf{C}_2\left(M_H^2,0,p_2^2;m_f^2,m_f^2,m_f^2\right)
+3 \mathbf{C}_1\left(M_H^2,0,p_2^2;m_f^2,m_f^2,m_f^2\right)
\Big\}.
\end{eqnarray}
Applying the same procedure, the form factors
calculating from $W$ boson loop diagrams
are expressed as follows
\begin{eqnarray}
\mathcal{A}_{00}^{(W)} &=&
M_W^2 
\Big\{
\mathbf{B}_0\left(p_2^2;M_W^2,M_W^2\right)
+8 \left[p_2^2-M_H^2\right]
\mathbf{C}_0\left(0,p_2^2,M_H^2;M_W^2,M_W^2,M_W^2\right)
\Big\}
\nonumber\\
&&
-\left[M_H^2+2 (d-1) M_W^2\right]
\Big[
\mathbf{B}_0\left(M_H^2;M_W^2,M_W^2\right)
-4 
\mathbf{C}_{00}\left(0,p_2^2,M_H^2;M_W^2,M_W^2,M_W^2\right)
\Big]
\nonumber\\
&&
-M_W^2 \mathbf{B}_0(0;M_W^2,M_W^2)
\\
\mathcal{A}_{21}^{(W)}&=&
2M_W^2 \Big\{
4 (d-1) \Big[ 
\mathbf{C}_{22}\left(0,p_2^2,M_H^2;M_W^2,M_W^2,M_W^2\right)
+ \mathbf{C}_{12}\left(0,p_2^2,M_H^2;M_W^2,M_W^2,M_W^2\right)
\Big]
\nonumber\\
&&
+2(2d+1) \mathbf{C}_2\left(0,p_2^2,M_H^2;M_W^2,M_W^2,M_W^2\right)
+11 \mathbf{C}_0\left(0,p_2^2,M_H^2;M_W^2,M_W^2,M_W^2\right)
\nonumber\\
&&
+3 \mathbf{C}_1\left(0,p_2^2,M_H^2;M_W^2,M_W^2,M_W^2\right)
\Big\}
+4 M_H^2 
\Big\{
\mathbf{C}_{22}\left(0,p_2^2,M_H^2;M_W^2,M_W^2,M_W^2\right)
\nonumber\\
&&
+\mathbf{C}_{12}\left(0,p_2^2,M_H^2;M_W^2,M_W^2,M_W^2\right)
+\mathbf{C}_2\left(0,p_2^2,M_H^2;M_W^2,M_W^2,M_W^2\right)
\Big\},
\\
\mathcal{A}_{22}^{(W)} &=&
4 \left[M_H^2+2 (d-1) M_W^2\right] 
\mathbf{C}_{22}\left(0,p_2^2,M_H^2;M_W^2,M_W^2,M_W^2\right)
\nonumber\\
&&
+2 \left[M_H^2+(2 d+3) M_W^2\right] 
\mathbf{C}_2\left(0,p_2^2,M_H^2;M_W^2,M_W^2,M_W^2\right)
\nonumber\\
&&
+4 M_W^2 \mathbf{C}_0\left(0,p_2^2,M_H^2;M_W^2,M_W^2,M_W^2\right).
\end{eqnarray}
Further, one-loop form factors for this channel 
with contributing of charged scalars in the loop 
diagrams are obtained
\begin{eqnarray}
\mathcal{A}^{(S_i)}_{00} &=&
-2 \Big\{
\mathbf{B}_0\left(M_H^2;M_{S_i}^2,M_{S_i}^2\right)-4
\mathbf{C}_{00}\left(M_H^2,0,p_2^2;M_{S_i}^2,M_{S_i}^2,M_{S_i}^2\right)
\Big\}, \\
\mathcal{A}^{(S_i)}_{21} &=&
8\Big\{
\mathbf{C}_{12}\left(M_H^2,0,p_2^2;M_{S_i}^2,M_{S_i}^2,M_{S_i}^2\right)
+\mathbf{C}_{11}\left(M_H^2,0,p_2^2;M_{S_i}^2,M_{S_i}^2,M_{S_i}^2\right)
\nonumber\\
&&
+\mathbf{C}_1\left(M_H^2,0,p_2^2;M_{S_i}^2,M_{S_i}^2,M_{S_i}^2\right)
\Big\},
\\
\mathcal{A}^{(S_i)}_{22} &=&
4\Big\{
2 \mathbf{C}_{22}\left(M_H^2,0,p_2^2;M_{S_i}^2,M_{S_i}^2,M_{S_i}^2\right)
+4 \mathbf{C}_{12}\left(M_H^2,0,p_2^2;M_{S_i}^2,M_{S_i}^2,M_{S_i}^2\right)
\nonumber\\
&&
+2 \mathbf{C}_{11}\left(M_H^2,0,p_2^2;M_{S_i}^2,M_{S_i}^2,M_{S_i}^2\right)
+\mathbf{C}_0\left(M_H^2,0,p_2^2;M_{S_i}^2,M_{S_i}^2,M_{S_i}^2\right)
\nonumber\\
&&
+3 \mathbf{C}_2\left(M_H^2,0,p_2^2;M_{S_i}^2,M_{S_i}^2,M_{S_i}^2\right)
+3 \mathbf{C}_1\left(M_H^2,0,p_2^2;M_{S_i}^2,M_{S_i}^2,M_{S_i}^2\right)
\Big\}.   
\end{eqnarray}
We find that all form factors in this subsection 
can be obtained by taking $p_1^2\rightarrow0$ from the results in two off-shell
photons.

\section*{Appendix $B$: Feynman rules for 
$H\rightarrow \gamma\gamma$ in $R_{\xi }$ gauge}
Feynman rules for 
$H\rightarrow \gamma\gamma$ in $R_{\xi}$ gauge devoted in this appendix. 
\begin{center}
\begin{table}[h!]
\centering
\scalebox{1.0}
{\begin{tabular}{l@{\hspace{2cm}}l }
\hline \hline
\textbf{Particle types} & \textbf{Propagators}\\
\hline \hline \\
Fermions & $\dfrac{i(\slashed{p} + m_f)}{p^2-m_f^2}$\\
$W$ boson 
& $\dfrac{- i}{p^2 - M_W^2} 
\Bigg[ g^{\mu \nu} - (1 - \xi) \dfrac{p^\mu p^\nu}{p^2 - M_{\xi}^2}  \Bigg]$ \\
Goldstone boson  
& $\dfrac{i}{p^2 - M_{\xi}^2}$ \\
Ghost 
& $\dfrac{i}{p^2 - M_{\xi}^2}$ \\
Charged scalar 
& $\dfrac{i}{p^2 -  M_{S_i}^2}$ \\ \\
\hline \hline \\
\end{tabular}}
\caption{Feynman rules involving the 
decay $H \rightarrow \gamma \gamma$ through 
fermion and $W$ boson, charged scalar loop diagrams
in the $R_{\xi}$ gauge.
\label{Feynman rules table}}
\end{table}
\end{center}

\begin{table}[h!]
\begin{center}
\begin{tabular}{l@{\hspace{2cm}}l} 
\hline \hline
\textbf{Vertices} & \textbf{Couplings}   \\ \hline \hline
\\
$A_\mu f\bar{f}$ & $i e Q_f\gamma_{\mu}$ \\
$H f\bar{f}$ & $-i gm_f / (2M_W)$ \\
$H \cdot W_\mu \cdot W_\nu $ 
& $i g M_W g_{\mu \nu}$  \\
$A_\mu (p_1) \cdot W^+_\nu (p_2) \cdot W^-_\lambda (p_3)$
& $-i e \Gamma_{\mu \nu \lambda} (p_1, p_2, p_3)$ \\
$A_\mu \cdot A_\nu \cdot W^+_\alpha \cdot W^-_\beta$
& $-i e^2 S_{\mu \nu, \alpha \beta}$
\\
$H (p_1) \cdot W_\mu \cdot \chi (p_2)$
& $-i \dfrac{g}{2} (p_2 - p_1)_\mu$ \\
$A_\mu \cdot W_\nu \cdot \chi$
& $-i e M_W g_{\mu \nu}$ \\
$H \cdot \chi \cdot \chi$
& $-i g M_H^2 / (2 M_W)$ \\
$A_\mu \cdot \chi (p_1) \cdot \chi (p_2)$
& $-i e (p_2 - p_1)_\mu$ \\
$A_\mu \cdot A_\nu \cdot \chi \cdot \chi$
& $i 2 e^2 g_{\mu \nu}$ \\
$H \cdot A_\mu \cdot W_\nu \cdot \chi$
& $-i e \dfrac{g}{2} g_{\mu \nu}$ \\
$H \cdot c \cdot c$
& $-i \xi \dfrac{g}{2}  M_W$ \\
$A_\mu \cdot c \cdot c$
& $-i e p_\mu$ \\
$HS_i\bar{S}_i$
& $i \lambda_{HS_i\bar{S}_i}$
\\
$A_{\mu}S_i(q_1)\bar{S}_i(q_2)$
& $i e Q_{S_i} (q_2-q_1)_{\mu}$\\
$A_{\mu}A_{\nu}S_i\bar{S}_i$
& $2i e^2 Q_{S_i}^2g_{\mu\nu}$ \\
\\ \hline \hline
\end{tabular}
\end{center}
\caption{Couplings involving 
the decay $H \rightarrow \gamma \gamma$ 
through fermion 
and $W$ boson, charged scalar loops 
in a in $R_{\xi}$
gauge with the short notation for the 
standard Lorentz tensors of the gauge boson self couplings 
$\Gamma_{\mu \nu \lambda} (p_1, p_2, p_3) 
= g_{\mu \nu} (p_1 - p_2)_\lambda + g_{\lambda \nu}
(p_2 - p_3)_\mu + g_{\mu \lambda} (p_3 - p_1)_\nu$ 
and $S_{\mu \nu, \alpha \beta} = 2 g_{\mu \nu} g_{\alpha \beta} 
- g_{\mu \alpha} g_{\nu \beta} - g_{\mu \beta} g_{\nu \alpha}$.
\label{couplings table}}
\end{table}
\section*{Appendix $C$: Amplitude 
$H\rightarrow  \gamma \gamma$ in $R_{\xi}$
gauge} 
One-loop Feynman amplitudes for the process 
$H\rightarrow  \gamma \gamma$ in $R_{\xi}$
gauge are shown in this appendix. For fermion
loop diagrams, one has
\begin{eqnarray}
 \mathcal{A}^{(\text{fermion})}_{(1+2)} &=&
\dfrac{g m_f Q_f^2e^2 }{M_W}
\int
\dfrac{{\rm d}^d k}{(2 \pi)^d}
\,
\dfrac{\text{Tr}
\{
(\slashed{k} + m_f) \gamma^\mu (\slashed{k} 
- \slashed{p_1} + m_f) \gamma^\nu (\slashed{k} - \slashed{p} + m_f)
\}
}{
(k^2-m_f^2) [(k-p_1)^2-m_f^2] [(k-p)^2-m_f^2]}
\varepsilon_{\mu}^* (p_1)
\varepsilon_{\nu}^*(p_2). 
\nonumber\\
\end{eqnarray}

\newpage
We next show Feynman amplitude for $W$ boson loop
diagrams

\subsection*{ \underline{Diagram $a$}}%
The Feynman amplitude for $W$ boson loop
diagrams are decomposed into $8$ terms as follows:
\begin{eqnarray}
\mathcal{A}^{(a)}(\xi)
&=&
\mathcal{A}_{111}(\xi)
+ \mathcal{A}_{112}(\xi)
+ \mathcal{A}_{121}(\xi)
+ \mathcal{A}_{211}(\xi)
\n \\
&&
+ \mathcal{A}_{122}(\xi)
+ \mathcal{A}_{212}(\xi)
+ \mathcal{A}_{221}(\xi)
+ \mathcal{A}_{222}(\xi).
\end{eqnarray}
Where $\mathcal{A}_{111}(\xi), \cdots,\mathcal{A}_{222}(\xi)$ 
are corresponding to which term in the right hand side of
Eq.~(\ref{WProb}) is taken. These terms are written
\begin{eqnarray}
\mathcal{A}_{111}(\xi)
&=&
2 e^2 g M_W
\int
\dfrac{{\rm d}^d k}{(2 \pi)^d}
\,
g_{\alpha \beta}
\Gamma_{\mu \tau \lambda} (-p_1, k, -k + p_1)
\Gamma_{\nu \rho \delta} (-p_2, k-p_1, -k+p)
\,
\epsilon_\mu^*(p_1)
 \epsilon_\nu^*(p_2)
\n \\
&&
\times
\dfrac{g^{\alpha \tau} - k^\alpha k^\tau/M_W^2}{k^2 - M_W^2} 
\dfrac{g^{\lambda \rho} - (k-p_1)^\lambda (k-p_1)^\rho/M_W^2}{(k-p_1)^2 - M_W^2} 
\dfrac{g^{\beta \delta} - (k-p)^\beta (k-p)^\delta/M_W^2}{(k-p)^2 - M_W^2} 
,
\\ 
\n \\
\mathcal{A}_{112}(\xi)
&=&
2 e^2 g M_W
\int
\dfrac{{\rm d}^d k}{(2 \pi)^d}
\,
g_{\alpha \beta}
\Gamma_{\mu \tau \lambda} (-p_1, k, -k + p_1)
\Gamma_{\nu \rho \delta} (-p_2, k-p_1, -k+p)
\,
\epsilon_\mu^*(p_1)
 \epsilon_\nu^*(p_2)
\n \\
&&
\times
\dfrac{g^{\alpha \tau} - k^\alpha k^\tau/M_W^2}{k^2 - M_W^2} 
\dfrac{g^{\lambda \rho} - (k-p_1)^\lambda (k-p_1)^\rho/M_W^2}{(k-p_1)^2 - M_W^2} 
\dfrac{(k-p)^\beta (k-p)^\delta/M_W^2}{(k-p)^2 - M_{\xi}^2} 
,
\end{eqnarray}

\begin{eqnarray}
\mathcal{A}_{121}(\xi)
&=&
2 e^2 g M_W
\int
\dfrac{{\rm d}^d k}{(2 \pi)^d}
\,
g_{\alpha \beta}
\Gamma_{\mu \tau \lambda} (-p_1, k, -k + p_1)
\Gamma_{\nu \rho \delta} (-p_2, k-p_1, -k+p)
\,
\epsilon_\mu^*(p_1)
 \epsilon_\nu^*(p_2)
\n \\
&&
\times
\dfrac{g^{\alpha \tau} - k^\alpha k^\tau/M_W^2}{k^2 - M_W^2} 
\dfrac{(k-p_1)^\lambda (k-p_1)^\rho/M_W^2}{(k-p_1)^2 - M_{\xi}^2} 
\dfrac{g^{\beta \delta} - (k-p)^\beta (k-p)^\delta/M_W^2}{(k-p)^2 - M_W^2} 
,
\\
\n \\
\mathcal{A}_{211}(\xi)
&=&
2 e^2 g M_W
\int
\dfrac{{\rm d}^d k}{(2 \pi)^d}
\,
g_{\alpha \beta}
\Gamma_{\mu \tau \lambda} (-p_1, k, -k + p_1)
\Gamma_{\nu \rho \delta} (-p_2, k-p_1, -k+p)
\,
\epsilon_\mu^*(p_1)
 \epsilon_\nu^*(p_2)
\n \\
&&
\times
\dfrac{k^\alpha k^\tau/M_W^2}{k^2 - M_{\xi}^2} 
\dfrac{g^{\lambda \rho} - (k-p_1)^\lambda (k-p_1)^\rho/M_W^2}{(k-p_1)^2 - M_W^2} 
\dfrac{g^{\beta \delta} - (k-p)^\beta (k-p)^\delta/M_W^2}{(k-p)^2 - M_W^2} 
,
\\
\n \\
\mathcal{A}_{122}(\xi)
&=&
2 e^2 g M_W
\int
\dfrac{{\rm d}^d k}{(2 \pi)^d}
\,
g_{\alpha \beta}
\Gamma_{\mu \tau \lambda} (-p_1, k, -k + p_1)
\Gamma_{\nu \rho \delta} (-p_2, k-p_1, -k+p)
\,
\epsilon_\mu^*(p_1)
 \epsilon_\nu^*(p_2)
\n \\
&&
\times
\dfrac{g^{\alpha \tau} - k^\alpha k^\tau/M_W^2}{k^2 - M_W^2} 
\dfrac{(k-p_1)^\lambda (k-p_1)^\rho/M_W^2}{(k-p_1)^2 - M_{\xi}^2} 
\dfrac{(k-p)^\beta (k-p)^\delta/M_W^2}{(k-p)^2 - M_{\xi}^2} 
,
\\
\n \\
\mathcal{A}_{212}(\xi)
&=&
2 e^2 g M_W
\int
\dfrac{{\rm d}^d k}{(2 \pi)^d}
\,
g_{\alpha \beta}
\Gamma_{\mu \tau \lambda} (-p_1, k, -k + p_1)
\Gamma_{\nu \rho \delta} (-p_2, k-p_1, -k+p)
\,
\epsilon_\mu^*(p_1)
 \epsilon_\nu^*(p_2)
\n \\
&&
\times
\dfrac{k^\alpha k^\tau/M_W^2}{k^2 - M_{\xi}^2} 
\dfrac{g^{\lambda \rho} - (k-p_1)^\lambda (k-p_1)^\rho/M_W^2}{(k-p_1)^2 - M_W^2} 
\dfrac{(k-p)^\beta (k-p)^\delta/M_W^2}{(k-p)^2 - M_{\xi}^2} 
,
\\
\n \\
\mathcal{A}_{221}(\xi)
&=&
2 e^2 g M_W
\int
\dfrac{{\rm d}^d k}{(2 \pi)^d}
\,
g_{\alpha \beta}
\Gamma_{\mu \tau \lambda} (-p_1, k, -k + p_1)
\Gamma_{\nu \rho \delta} (-p_2, k-p_1, -k+p)
\,
\epsilon_\mu^*(p_1)
 \epsilon_\nu^*(p_2)
\n \\
&&
\times
\dfrac{k^\alpha k^\tau/M_W^2}{k^2 - M_{\xi}^2} 
\dfrac{(k-p_1)^\lambda (k-p_1)^\rho/M_W^2}{(k-p_1)^2 - M_{\xi}^2} 
\dfrac{g^{\beta \delta} - (k-p)^\beta (k-p)^\delta/M_W^2}{(k-p)^2 - M_W^2} 
,
\\
\n \\
\mathcal{A}_{222}(\xi)
&=&
2 e^2 g M_W
\int
\dfrac{{\rm d}^d k}{(2 \pi)^d}
\,
g_{\alpha \beta}
\Gamma_{\mu \tau \lambda} (-p_1, k, -k + p_1)
\Gamma_{\nu \rho \delta} (-p_2, k-p_1, -k+p)
\,
\epsilon_\mu^*(p_1)
 \epsilon_\nu^*(p_2)
\n \\
&&
\times
\dfrac{k^\alpha k^\tau/M_W^2}{k^2 - M_{\xi}^2} 
\dfrac{(k-p_1)^\lambda (k-p_1)^\rho/M_W^2}{(k-p_1)^2 - M_{\xi}^2} 
\dfrac{(k-p)^\beta (k-p)^\delta/M_W^2}{(k-p)^2 - M_{\xi}^2} 
.
\end{eqnarray}

\subsection*{ \underline{Diagram $b$}}%

\begin{eqnarray}
\mathcal{A}^{(b)}(\xi)
&=& 
\mathcal{A}_{11}(\xi)
+ \mathcal{A}_{12}(\xi)
+ \mathcal{A}_{21}(\xi)
+ \mathcal{A}_{22}(\xi)
\end{eqnarray}
where
\begin{eqnarray}
\mathcal{A}_{11}(\xi)
&=&
-e^2 g M_W
\int
\dfrac{{\rm d}^d k}{(2 \pi)^d}
\,
g_{\alpha \beta}
S_{\mu \nu, \lambda \rho}
\,
\epsilon_\mu^*(p_1)
 \epsilon_\nu^*(p_2)
\n \\
&& \times
\dfrac{g^{\alpha \lambda} - k^\alpha k^\lambda/M_W^2}{k^2 - M_W^2} 
\dfrac{g^{\beta \rho} - (k-p)^\beta (k-p)^\rho/M_W^2}{(k-p)^2 - M_W^2} 
,
\\
\n \\
\mathcal{A}_{12}(\xi)
&=&
-e^2 g M_W
\int
\dfrac{{\rm d}^d k}{(2 \pi)^d}
\,
g_{\alpha \beta}
S_{\mu \nu, \lambda \rho}
\,
\epsilon_\mu^*(p_1)
 \epsilon_\nu^*(p_2)
\n \\
&& \times
\dfrac{g^{\alpha \lambda} - k^\alpha k^\lambda/M_W^2}{k^2 - M_W^2} 
\dfrac{(k-p)^\beta (k-p)^\rho/M_W^2}{(k-p)^2 - M_{\xi}^2} 
,
\end{eqnarray}

\begin{eqnarray}
\mathcal{A}_{21}(\xi)
&=&
-e^2 g M_W
\int
\dfrac{{\rm d}^d k}{(2 \pi)^d}
\,
g_{\alpha \beta}
S_{\mu \nu, \lambda \rho}
\,
\epsilon_\mu^*(p_1)
 \epsilon_\nu^*(p_2)
\n \\
&& \times
\dfrac{k^\alpha k^\lambda/M_W^2}{k^2 - M_{\xi}^2} 
\dfrac{g^{\beta \rho} - (k-p)^\beta (k-p)^\rho/M_W^2}{(k-p)^2 - M_W^2} 
,
\\
\n \\
\mathcal{A}_{22}(\xi)
&=&
-e^2 g M_W
\int
\dfrac{{\rm d}^d k}{(2 \pi)^d}
\,
g_{\alpha \beta}
S_{\mu \nu, \lambda \rho}
\,
\epsilon_\mu^*(p_1)
 \epsilon_\nu^*(p_2)
\n \\
&& \times
\dfrac{k^\alpha k^\lambda/M_W^2}{k^2 - M_{\xi}^2} 
\dfrac{(k-p)^\beta (k-p)^\rho/M_W^2}{(k-p)^2 - M_{\xi}^2} 
.
\end{eqnarray}

\subsection*{ \underline{Diagram $c$}}%

\begin{eqnarray}
\mathcal{A}^{(c)}(\xi)
&=& 
\mathcal{A}_{110}(\xi)
+ \mathcal{A}_{120}(\xi)
+ \mathcal{A}_{210}(\xi)
+ \mathcal{A}_{220}(\xi)
\end{eqnarray}
where
\begin{eqnarray}
\mathcal{A}_{110}(\xi)
&=&
2 e^2 g M_W
\int
\dfrac{{\rm d}^d k}{(2 \pi)^d}
\,
(k-2p)_\alpha
g_{\nu \rho}
\Gamma_{\mu \tau \lambda} (-p_1, k, -k+p_1)
\,
\epsilon_\mu^*(p_1)
 \epsilon_\nu^*(p_2)
\n \\
&&
\times
\dfrac{g^{\alpha \tau} - k^\alpha k^\tau/M_W^2}{k^2 - M_W^2} 
\dfrac{g^{\lambda \rho} - (k-p_1)^\lambda (k-p_1)^\rho/M_W^2}{(k-p_1)^2 - M_W^2} 
\dfrac{1}{(k-p)^2 - M_{\xi}^2}
,
\\
\n \\
\mathcal{A}_{120}(\xi)
&=&
2 e^2 g M_W
\int
\dfrac{{\rm d}^d k}{(2 \pi)^d}
\,
(k-2p)_\alpha
g_{\nu \rho}
\Gamma_{\mu \tau \lambda} (-p_1, k, -k+p_1)
\,
\epsilon_\mu^*(p_1)
 \epsilon_\nu^*(p_2)
\n \\
&&
\times
\dfrac{g^{\alpha \tau} - k^\alpha k^\tau/M_W^2}{k^2 - M_W^2} 
\dfrac{(k-p_1)^\lambda (k-p_1)^\rho/M_W^2}{(k-p_1)^2 - M_{\xi}^2} 
\dfrac{1}{(k-p)^2 - M_{\xi}^2}
,
\\
\n \\
\mathcal{A}_{210}(\xi)
&=&
2 e^2 g M_W
\int
\dfrac{{\rm d}^d k}{(2 \pi)^d}
\,
(k-2p)_\alpha
g_{\nu \rho}
\Gamma_{\mu \tau \lambda} (-p_1, k, -k+p_1)
\,
\epsilon_\mu^*(p_1)
 \epsilon_\nu^*(p_2)
\n \\
&&
\times
\dfrac{k^\alpha k^\tau/M_W^2}{k^2 - M_{\xi}^2} 
\dfrac{g^{\lambda \rho} - (k-p_1)^\lambda (k-p_1)^\rho/M_W^2}{(k-p_1)^2 - M_W^2} 
\dfrac{1}{(k-p)^2 - M_{\xi}^2}
,
\\
\n \\
\mathcal{A}_{220}(\xi)
&=&
2 e^2 g M_W
\int
\dfrac{{\rm d}^d k}{(2 \pi)^d}
\,
(k-2p)_\alpha
g_{\nu \rho}
\Gamma_{\mu \tau \lambda} (-p_1, k, -k+p_1)
\,
\epsilon_\mu^*(p_1)
 \epsilon_\nu^*(p_2)
\n \\
&&
\times
\dfrac{k^\alpha k^\tau/M_W^2}{k^2 - M_{\xi}^2} 
\dfrac{(k-p_1)^\lambda (k-p_1)^\rho/M_W^2}{(k-p_1)^2 - M_{\xi}^2} 
\dfrac{1}{(k-p)^2 - M_{\xi}^2}
.
\end{eqnarray}

\subsection*{ \underline{Diagram $d$}}%

\begin{eqnarray}
\mathcal{A}^{(d)}(\xi)
&=& 
\mathcal{A}_{10}(\xi)
+ \mathcal{A}_{20}(\xi)
\end{eqnarray}
where
\begin{eqnarray}
\mathcal{A}_{10}(\xi)
&=&
-2 e^2 g M_W
\int
\dfrac{{\rm d}^d k}{(2 \pi)^d}
\,
g_{\mu \lambda}
g_{\nu \rho}
\dfrac{g^{\lambda \rho} - (k-p_1)^\lambda (k-p_1)^\rho/M_W^2}{(k-p_1)^2 - M_W^2} 
\dfrac{\epsilon_\mu^*(p_1)
 \epsilon_\nu^*(p_2)}{(k-p)^2 - M_{\xi}^2}
,
\\
\n \\
\mathcal{A}_{20}(\xi)
&=&
-2 e^2 g M_W
\int
\dfrac{{\rm d}^d k}{(2 \pi)^d}
\,
g_{\mu \lambda}
g_{\nu \rho}
\dfrac{(k-p_1)^\lambda (k-p_1)^\rho/M_W^2}{(k-p_1)^2 - M_{\xi}^2} 
\dfrac{\epsilon_\mu^*(p_1)
 \epsilon_\nu^*(p_2)}{(k-p)^2 - M_{\xi}^2}
.
\end{eqnarray}

\subsection*{ \underline{Diagram $e$}}%

\begin{eqnarray}
\mathcal{A}^{(e)}(\xi)
&=& 
\mathcal{A}_{100}(\xi)
+ \mathcal{A}_{200}(\xi)
\end{eqnarray}
where
\begin{eqnarray}
\mathcal{A}_{100}(\xi)
&=&
-2 e^2 g M_W
\int
\dfrac{{\rm d}^d k}{(2 \pi)^d}
\,
g_{\mu \tau}
(k-2p)_\alpha
(-2k+2p_1+p_2)_\nu
\n \\
&&
\times
\dfrac{g^{\alpha \tau} - k^\alpha k^\tau/M_W^2}{k^2 - M_W^2} 
\dfrac{1}{(k-p_1)^2 - M_{\xi}^2}
\dfrac{1}{(k-p)^2 - M_{\xi}^2}
\,
\epsilon_\mu^*(p_1)
 \epsilon_\nu^*(p_2)
,
\\
\n \\
\mathcal{A}_{200}(\xi)
&=&
-2 e^2 g M_W
\int
\dfrac{{\rm d}^d k}{(2 \pi)^d}
\,
g_{\mu \tau}
(k-2p)_\alpha
(-2k+2p_1+p_2)_\nu
\n \\
&&
\times
\dfrac{k^\alpha k^\tau/M_W^2}{k^2 - M_{\xi}^2} 
\dfrac{1}{(k-p_1)^2 - M_{\xi}^2}
\dfrac{1}{(k-p)^2 - M_{\xi}^2}
\,
\epsilon_\mu^*(p_1)
 \epsilon_\nu^*(p_2)
.
\end{eqnarray}

\subsection*{ \underline{Diagram $f$}}%

\begin{eqnarray}
\mathcal{A}^{(f)}(\xi)
&=& 
\mathcal{A}_{101}(\xi)
+ \mathcal{A}_{102}(\xi)
+ \mathcal{A}_{201}(\xi)
+ \mathcal{A}_{202}(\xi)
\end{eqnarray}
where
\begin{eqnarray}
\mathcal{A}_{101}(\xi)
&=&
-2 e^2 g M_W^3
\int
\dfrac{{\rm d}^d k}{(2 \pi)^d}
\,
g_{\alpha \beta}
g_{\mu \tau}
g_{\nu \delta}
\,
\epsilon_\mu^*(p_1)
 \epsilon_\nu^*(p_2)
\n \\
&&
\times
\dfrac{g^{\alpha \tau} - k^\alpha k^\tau/M_W^2}{k^2 - M_W^2} 
\dfrac{1}{(k-p_1)^2 - M_{\xi}^2}
\dfrac{g^{\beta \delta} - (k-p)^\beta (k-p)^\delta/M_W^2}{(k-p)^2 - M_W^2} 
,
\\
\n \\
\mathcal{A}_{102}(\xi)
&=&
-2 e^2 g M_W^3
\int
\dfrac{{\rm d}^d k}{(2 \pi)^d}
\,
g_{\alpha \beta}
g_{\mu \tau}
g_{\nu \delta}
\,
\epsilon_\mu^*(p_1)
 \epsilon_\nu^*(p_2)
\n \\
&&
\times
\dfrac{g^{\alpha \tau} - k^\alpha k^\tau/M_W^2}{k^2 - M_W^2} 
\dfrac{1}{(k-p_1)^2 - M_{\xi}^2}
\dfrac{(k-p)^\beta (k-p)^\delta/M_W^2}{(k-p)^2 - M_{\xi}^2} 
,
\\
\n \\
\mathcal{A}_{201}(\xi)
&=&
-2 e^2 g M_W^3
\int
\dfrac{{\rm d}^d k}{(2 \pi)^d}
\,
g_{\alpha \beta}
g_{\mu \tau}
g_{\nu \delta}
\,
\epsilon_\mu^*(p_1)
 \epsilon_\nu^*(p_2)
\n \\
&&
\times
\dfrac{k^\alpha k^\tau/M_W^2}{k^2 - M_{\xi}^2} 
\dfrac{1}{(k-p_1)^2 - M_{\xi}^2}
\dfrac{g^{\beta \delta} - (k-p)^\beta (k-p)^\delta/M_W^2}{(k-p)^2 - M_W^2} 
,
\\
\n \\
\mathcal{A}_{202}(\xi)
&=&
-2 e^2 g M_W^3
\int
\dfrac{{\rm d}^d k}{(2 \pi)^d}
\,
g_{\alpha \beta}
g_{\mu \tau}
g_{\nu \delta}
\,
\epsilon_\mu^*(p_1)
 \epsilon_\nu^*(p_2)
\n \\
&&
\times
\dfrac{k^\alpha k^\tau/M_W^2}{k^2 - M_{\xi}^2} 
\dfrac{1}{(k-p_1)^2 - M_{\xi}^2}
\dfrac{(k-p)^\beta (k-p)^\delta/M_W^2}{(k-p)^2 - M_{\xi}^2} 
.
\end{eqnarray}

\subsection*{ \underline{Diagram $g$}}%

\begin{eqnarray}
\mathcal{A}^{(g)}(\xi)
&=& 
\mathcal{A}_{010}(\xi)
+ \mathcal{A}_{020}(\xi)
\end{eqnarray}
where
\begin{eqnarray}
\mathcal{A}_{010}(\xi)
&=&
-e^2 g M_H^2 M_W
\int
\dfrac{{\rm d}^d k}{(2 \pi)^d}
\,
g_{\mu \lambda}
g_{\nu \rho}
\,
\epsilon_\mu^*(p_1)
 \epsilon_\nu^*(p_2)
\n \\
&&
\times
\dfrac{1}{k^2 - M_{\xi}^2}
\dfrac{g^{\lambda \rho} - (k-p_1)^\lambda (k-p_1)^\rho/M_W^2}{(k-p_1)^2 - M_W^2} 
\dfrac{1}{(k-p)^2 - M_{\xi}^2}
,
\\
\n \\
\mathcal{A}_{020}(\xi)
&=&
-e^2 g M_H^2 M_W
\int
\dfrac{{\rm d}^d k}{(2 \pi)^d}
\,
g_{\mu \lambda}
g_{\nu \rho}
\,
\epsilon_\mu^*(p_1)
 \epsilon_\nu^*(p_2)
\n \\
&&
\times
\dfrac{1}{k^2 - M_{\xi}^2}
\dfrac{(k-p_1)^\lambda (k-p_1)^\rho/M_W^2}{(k-p_1)^2 - M_{\xi}^2} 
\dfrac{1}{(k-p)^2 - M_{\xi}^2}
.
\end{eqnarray}

\subsection*{ \underline{Diagram $h$}}%

\begin{eqnarray}
\mathcal{A}^{(h)}(\xi)
&=& 
e^2 g \dfrac{M_H^2}{M_W}
\int
\dfrac{{\rm d}^d k}{(2 \pi)^d}
\,
(-2k+p_1)_\mu
(-2k+2p_1+p_2)_\nu
\n \\
&&
\times
\dfrac{1}{k^2 - M_{\xi}^2}
\dfrac{1}{(k-p_1)^2 - M_{\xi}^2}
\dfrac{1}{(k-p)^2 - M_{\xi}^2}
\,
\epsilon_\mu^*(p_1)
 \epsilon_\nu^*(p_2)
.
\end{eqnarray}

\subsection*{ \underline{Diagram $i$}}%

\begin{eqnarray}
\mathcal{A}^{(i)}(\xi)
&=& 
- e^2 g \dfrac{M_H^2}{M_W}
\int
\dfrac{{\rm d}^d k}{(2 \pi)^d}
\,
g_{\mu \nu}
\dfrac{1}{k^2 - M_{\xi}^2}
\dfrac{1}{(k-p)^2 - M_{\xi}^2}
\,
\epsilon_\mu^*(p_1)
 \epsilon_\nu^*(p_2)
 .
\end{eqnarray}

\subsection*{ \underline{Diagram $j$}}%

\begin{eqnarray}
\mathcal{A}^{(j)}(\xi)
&=& 
-2 e^2 g M_W
\,
\xi 
\int
\dfrac{{\rm d}^d k}{(2 \pi)^d}
\,
(k-p_1)_\mu
(k-p)_\nu
\n \\
&&
\times
\dfrac{1}{k^2 - M_{\xi}^2}
\dfrac{1}{(k-p_1)^2 - M_{\xi}^2}
\dfrac{1}{(k-p)^2 - M_{\xi}^2}
\,
\epsilon_\mu^*(p_1)
 \epsilon_\nu^*(p_2)
.
\end{eqnarray}

Feynman amplitude due to the charged scalar
particles exchanging in the loop diagrams
reads
\begin{eqnarray}
\mathcal{A}^{(S_i)}_{(1)}
&=& 
- 2 e^2 Q^2_{S_i} 
\lambda_{HS_i\bar{S}_i}
\int
\dfrac{{\rm d}^d k}{(2 \pi)^d}
\,
(-2k+p_1)_\mu
(-2k+2p_1+p_2)_\nu
\n \\
&&
\times
\dfrac{1}{k^2 -  M_{S_i}^2}
\dfrac{1}{(k-p_1)^2 -  M_{S_i}^2}
\dfrac{1}{(k-p)^2 - M_{S_i}^2}
\,
\epsilon_\mu^*(p_1)
 \epsilon_\nu^*(p_2)
,
\\
\n \\
\mathcal{A}^{(S_i)}_{(2)}
&=& 
2 e^2 Q_{S_i}^2
\lambda_{HS_i\bar{S}_i}
\int
\dfrac{{\rm d}^d k}{(2 \pi)^d}
\,
g_{\mu \nu}
\dfrac{1}{k^2 -  M_{S_i}^2}
\dfrac{1}{(k-p)^2 -  M_{S_i}^2}
\,
\epsilon_\mu^*(p_1)
 \epsilon_\nu^*(p_2)
 .
\end{eqnarray}

\end{document}